\documentclass[%
reprint,
superscriptaddress,
twocolumn,
amsmath,
amssymb,
aps,
pra,
]{revtex4-2}

\usepackage[
    colorlinks=true,
    linkcolor=blue,
    citecolor=blue,
    urlcolor=blue
]{hyperref}

\usepackage{dcolumn}
\usepackage{bm}
\usepackage{mathtools} 
\usepackage{color}  
\usepackage{braket}
\usepackage{tikz}
\usetikzlibrary{quantikz2}
\usepackage{float}
\usepackage[normalem]{ulem}
\usepackage{multirow}
\usepackage{siunitx}
\usetikzlibrary{snakes}
\usepackage{amsmath}
\usepackage{mhchem}
\usepackage{color,soul}
\usepackage[OT1, T1]{fontenc}
\DeclareTextSymbolDefault{\dh}{OT1}
\usepackage{graphicx}
\usepackage[caption=false]{subfig}
\usepackage[export]{adjustbox}
\usepackage{comment}

\usepackage{bm}
\usepackage{bbold}

\definecolor{mscolor}{rgb}{0,0.5,0.5}
\definecolor{akscolor}{rgb}{0.5,0,0}
\definecolor{tgcolor}{rgb}{0.2,0.8,0.2}

{}

\newcommand \be{\begin{equation}}
\newcommand \ee{\end{equation}}
\newcommand \bea{\begin{eqnarray}}
\newcommand \eea{\end{eqnarray}}
\newcommand \bse{\begin{subequations}}
\newcommand \ese{\end{subequations}}

\usepackage{xcolor} 
\newcommand {\rsub}[1]{\textcolor{black}{#1}}
 
\newcommand{\UWM}{Department of Physics, University of Wisconsin-Madison, 1150 University Avenue, Madison, WI, 53706, USA}

\def\afterfi#1#2\fi{\fi#1}

\def\pflA#1#2{\dot#1\ifx.#2\else\afterfi{\pflB#2}\fi}
\def\pflB#1#2{\ifx.#2\dot#1\else#1\afterfi{\pflB#2}\fi}

\def\padotsA#1#2{\dot#1\ifx.#2\else\afterfi{\padotsA#2}\fi}

\def\pmiA#1#2{\ifx.#2\dot#1\else\afterfi{\pmiB#1#2}\fi}
\def\pmiB#1.{\overline{#1}}

\begin{document}

\title{ 
Efficient and compact quantum network node based on a parabolic mirror on an optical chip}
 
\author{A. Safari}
\email{asafari@wisc.edu}
\affiliation{\UWM}
\author{E. Oh}
\affiliation{\UWM}
\author{P. Huft}
\affiliation{\UWM}
\affiliation{Present address: QuEra Computing, 1380 Soldiers Field Road, Boston, MA 02135, USA}
\author{G. Chase}
\affiliation{\UWM}
\affiliation{Present address: Department of Physics and Nuclear Engineering, United States Military Academy, West Point, NY, 10996, USA}
\author{J. Zhang}
\affiliation{\UWM}
\affiliation{Present address: Max-Planck-Institut f\"ur Quantenoptik, 85748 Garching, Germany}
\author{M. Saffman}
\affiliation{\UWM}

\date{\today}

\begin{abstract}
We demonstrate a neutral atom networking node that combines high photon collection efficiency with high atom–photon entanglement fidelity in a compact, fiber-integrated platform. A parabolic mirror is used both to form the trap and to collect fluorescence from a single rubidium atom, intrinsically mode-matching $\sigma$ polarized emitted photons to the fiber and rendering the system largely insensitive to small imperfections or drifts. The core optics consist of millimeter-scale components that are pre-aligned, rigidly bonded on a monolithic in-vacuum assembly, and interfaced entirely via optical fibers. With this design, we measure an overall photon collection and detection efficiency of \rsub{$5\%$}, from which we infer an overall collection efficiency of \rsub{$9\%$} after the single-mode fiber coupling. We generate atom–photon entangled states with a raw Bell-state fidelity of 0.93 and an inferred fidelity of 0.98 after correcting for atom readout errors. The same node design has been realized in two independent setups with comparable performance and is compatible with adding high-NA objective lenses to create and control atomic arrays at each node. Our results establish a robust, cavity-free neutral atom interface that operates near the limit set by the collection optics numerical aperture and provides a practical building block for scalable quantum network nodes and repeaters.
\end{abstract}

\maketitle


\section{Introduction} 
Quantum information processing has progressed from proof-of-principle few-qubit demonstrations to programmable devices with hundreds or more physical qubits. However, scaling a single monolithic processor to the sizes required for fault-tolerant quantum computing or large-scale simulation poses severe architectural and engineering challenges. A complementary route is to adopt a modular architecture in which multiple smaller quantum processors and sensors are interconnected via photonic links. In such a setting, a quantum network \cite{Kimble2008, Reiserer2015, Wehner2018} provides not only a path to larger effective processor sizes, but also enables new capabilities such as blind and distributed quantum computation~\cite{Drmota_PRL_2024, Polacchi2023, Afzal2024, Main_Nature_2025}, entangled sensor and clock networks~\cite{Komar2014, Komar2016PRL, BACON_Nature_2021, Malia_Nature_2022, Nichol_Nature_2022, Gottesman2012, Marchese2023}, and device-independent quantum communication protocols~\cite{Liu2022PRL, Nadlinger2022, Zhang2022, Zapatero2023}.

A fundamental resource in a quantum network is atom-photon entanglement which enables remote entanglement of matter qubits, for example through interference of the photons and Bell-state measurements~\cite{SimonPRL2003, DuanPRL2003, BrownePRL2003, FengPRL2003}. Remote entanglement has been realized on several platforms, including trapped ions~\cite{Stephenson2020, Krutyanskiy2023, OReilly2025}, color centers~\cite{Pompili2021}, quantum dots~\cite{Stockill2017}, and neutral atoms~\cite{Welte2021, vanLeent2022}. The generation rate and fidelity of remote entanglement links are largely determined by the light–matter interface at each node and by the coherence properties of the stationary qubits.

\begin{figure*}[t!]
\includegraphics[width=\textwidth]{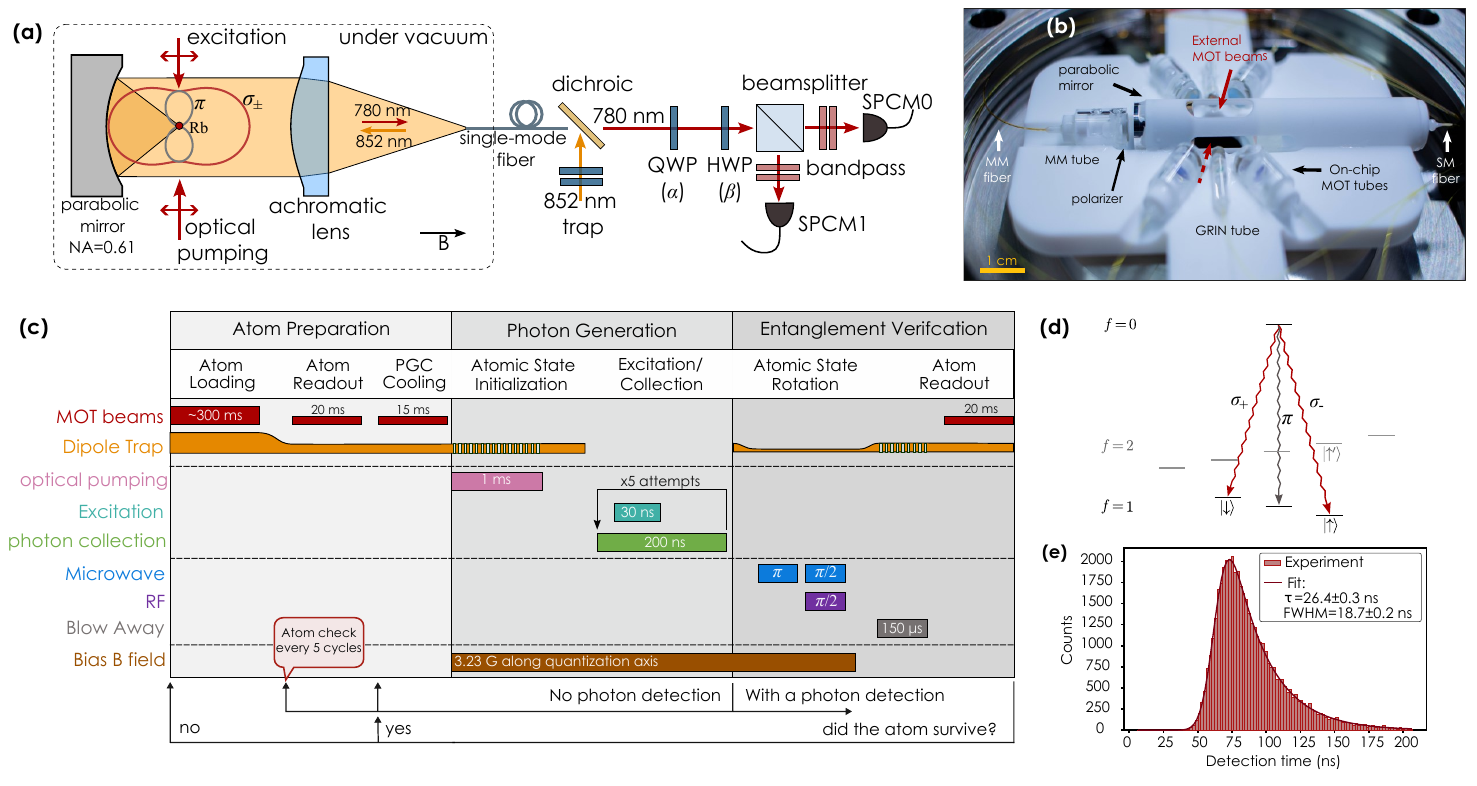}
\caption{Optical setup and experimental sequence. (a) Setup used for single-atom and single-photon characterization as well as atom-photon entanglement verification. For the $g^{(2)}$ measurement, the beamsplitter is a 50:50 non-polarizing beamsplitter. For atom-photon entanglement we replaced the beamsplitter with a polarizing beamsplitter (PBS). (b) Image of the in-vacuum components which include all the optics required at the core of the experiment, except two external MOT beams. (c) Experimental sequence for atom-photon entanglement generation and verification. (d) The atom is initially prepared in $\ket{f=1,m_f=0}$ by optical pumping, then excited to $\ket{f=0,m_f=0}$ by a Gaussian pulse. Emitted photons with $\sigma_{\pm}$ polarization are collected and coupled to the SM fiber. (e) Histogram of the single-photon detection time. The fit is a convolution of a Gaussian with FWHM of $18.7~\mathrm{ns}$ with an exponential with a decay time of $26.4~\mathrm{ns}$ in agreement with the lifetime of the excited state.}
 \label{fig.figure1}
\end{figure*}

Neutral atoms in optical traps have recently emerged as a leading candidate for large-scale quantum processors; arrays containing thousands of qubits have been demonstrated \cite{Norcia2024,Gyger2024,Manetsch2025}. At the same time, there is rapid progress on neutral atom networking architectures that integrate atomic registers with efficient photon-collection optics or optical cavities, dual-species implementations for separate data and communication qubits, and telecom-band interfaces~\cite{Young2022,BHu2025,Seubert2025,LLi2025, Covey2023}. Despite these advances, realizing high-rate and high-fidelity remote entanglement between neutral atom nodes remains an outstanding challenge. A primary limitation is the efficiency with which single photons can be collected and coupled into single-mode fibers. In free space, high-NA microscope objectives can intercept a sizable fraction of the dipole emission pattern, but the overall efficiency after mode-matching into a single-mode fiber is typically at the one percent level per emitted photon, even for very large NA. This limitation has motivated cavity-enhanced architectures and multiplexed protocols, which can in principle boost the rate, but add considerable experimental complexity and, in the case of nanophotonic structures, are difficult to reconcile with high-fidelity Rydberg operations \cite{Ocola2024}.

\rsub{Here we demonstrate an efficient neutral-atom quantum networking node based on a parabolic mirror that achieves both high single-photon collection efficiency and high atom-photon entanglement fidelity without the use of a cavity. Mirror-based approaches have previously been studied for efficient fluorescence collection from trapped ions~\cite{Maiwald2012, LLuo2009, Lindlein2007, Shu2010, Kim2011} and from neutral atoms with spherical mirrors~\cite{Roy2012}, though not with coupling into a single-mode fiber for quantum networking. Our approach is based on a novel design that uses a parabolic mirror for both trapping single atoms and collecting the scattered photons. This geometry, inspired by~\cite{Garcia2013}, uses the same single-mode fiber and the same optical path for delivering the trapping light and collecting the scattered photons. As a result, the collected field is naturally well matched to the single-mode fiber input, up to differences arising from the atomic emission pattern and optical aberrations.}

In addition, we built the core of the optomechanical assembly from millimeter-scale optics that are pre-aligned and rigidly bonded on a monolithic platform inside the vacuum chamber, with all optics interfaced to fibers that are accessible outside the vacuum. We show that the overall photon collection and detection efficiency after the single-mode fiber reaches \rsub{$5\%$}, significantly exceeding typical values obtained with large external high-NA objectives and approaching the theoretical limit set by the mirror’s numerical aperture. Using this node, we generate atom–photon entangled states with a raw Bell-state fidelity of $0.93\pm0.05$, and an inferred fidelity of $\approx 0.98$ after correcting for independently characterized atom state measurement errors. This compact, fiber-integrated architecture provides a robust and near-optimal light–matter interface for neutral atoms and constitutes a key step toward scalable, plug-and-play quantum network nodes and repeaters based on neutral atom processors.

\section{Single atom trapping} 
For single-atom trapping as well as photon collection we use a parabolic mirror, as shown in Fig.~\ref{fig.figure1}(a). The single-mode (SM), non-polarization-maintaining fiber carries the trap light at $852~\mathrm{nm}$ which is collimated by the achromatic lens and focused by the parabolic mirror with numerical aperture $\mathrm{NA} = 0.61$. We used a tapered fiber as a point detector to characterize the focal point and measured  waists ($1/e^2$ intensity radius) of $w_{x(y)}\simeq  0.77(0.66)~\mu\mathrm{m}$, which are close to the diffraction limit and can trap a single atom. An advantage of the parabolic mirror design is that it is highly achromatic; we measured an axial chromatic shift of less than $200~\mathrm{nm}$ between the focal points of $852~\mathrm{nm}$ and $780~\mathrm{nm}$ light. The scattered light from the trapped atom is collected by the same optics and coupled back to the same single-mode fiber. The time-reversal symmetry in this design provides excellent stability and insensitivity to slight misalignment in the optics and guarantees good coupling efficiency of the scattered photons into the SM fiber~\cite{Garcia2013}. See Appendix \ref{app.alignment} for further details of the alignment and characterization process of the optical setup. Details of the fabrication and alignment of the apparatus are provided in Appendices \ref{app.vacuum} - \ref{app.electronics}.

To enhance optical stability, we built the core of the setup from millimeter-scale optics, prealigned and glued on a Macor platform (Fig.~\ref{fig.figure1}(b)). The on-chip optics include the parabolic mirror module, four of the magneto-optical trapping (MOT) beams, plus two sets of beams made from GRIN lenses for optical pumping and excitation of the single atom. All of these optics, interfaced with optical fibers, were built in a tabletop clean-room environment and aligned to overlap with the focal point of the parabolic mirror. 

The dipole trap light should preferably have linear polarization in order to achieve a good polarization gradient cooling (PGC) with the trapped atom~\cite{YSChin2017}. However, the SM fiber does not maintain polarization if the fiber is moved or the environmental temperature changes. To control the polarization of the dipole trap light, we used a  half-waveplate (HWP) and quarter-waveplate (QWP) before coupling into the fiber (Fig.~\ref{fig.figure1}(a)). The parabolic mirror has an on-axis hole in the middle with a diameter of $0.5~\mathrm{mm}$ to pass a small portion of the trap light for polarization measurement. The hole covers a small portion of the solid angle and does not reduce the effective NA of the parabolic mirror significantly. A polarizer with transmission along the vertical axis is placed on the back of the parabolic mirror after which the light couples into a multi-mode fiber. To achieve a linearly polarized dipole trap at the atom, we maximize the power measured at the output of the multi-mode fiber by rotating the waveplates. Although maximizing transmission is not a very sensitive method for polarization alignment, it is sufficient to achieve an atom temperature below $20~\mu\mathrm{K}$ \rsub{(Appendix~\ref{app.atomTemp})}.

The in-vacuum assembly includes eight optical fibers to carry light in and out of the chamber. The fibers leave the vacuum assembly through teflon plugs in swageloks \cite{Abraham1998}. To form a MOT, two external beams are aligned outside of the chamber to pass through the focus of the parabolic mirror and overlap with the on-chip MOT beams at an angle of $35^\circ$ with respect to the normal to the Macor plate. All MOT beams have a waist of $0.5~\mathrm{mm}$. Overlapping the cold atomic cloud with the focus of the parabolic mirror  results in single-atom trapping which increases the count level on the single-photon counting modules (SPCMs). See Fig.~\ref{fig.figure1}(a) for the setup. Hence, the apparatus does not require a camera; we measure the single-atom signals with the SPCMs only. 

To load a single atom, we start with a trap power of a couple of mW, resulting in a trap depth of $\sim1.5~\mathrm{mK}$. After detecting an atom with the SPCMs, the trap is lowered adiabatically to $\sim500~\mu\mathrm{K}$ and remains at this level for  most of the experimental sequence (Fig.~\ref{fig.figure1}(c)). Raman scattering of the trap light at $852~\mathrm{nm}$ in the SM fiber results in $780~\mathrm{nm}$ photons propagating backwards towards the SPCMs, which increases the background level by about $2~\mathrm{kcounts/s}$ on each SPCM after passing through band-pass filters with FWHM of $2~\mathrm{nm}$ centered at $780~\mathrm{nm}$. We use an exposure time of  $20~\mathrm{ms}$ for each atom readout, during which we detect about $100$ photons from each SPCM. The contribution from the background during the readout is about $40$ counts on each SPCM. The detuning and power of the cooling light are adjusted such that the readout also provides PGC cooling on the atom. 

\begin{figure*}
\includegraphics[width=\textwidth]{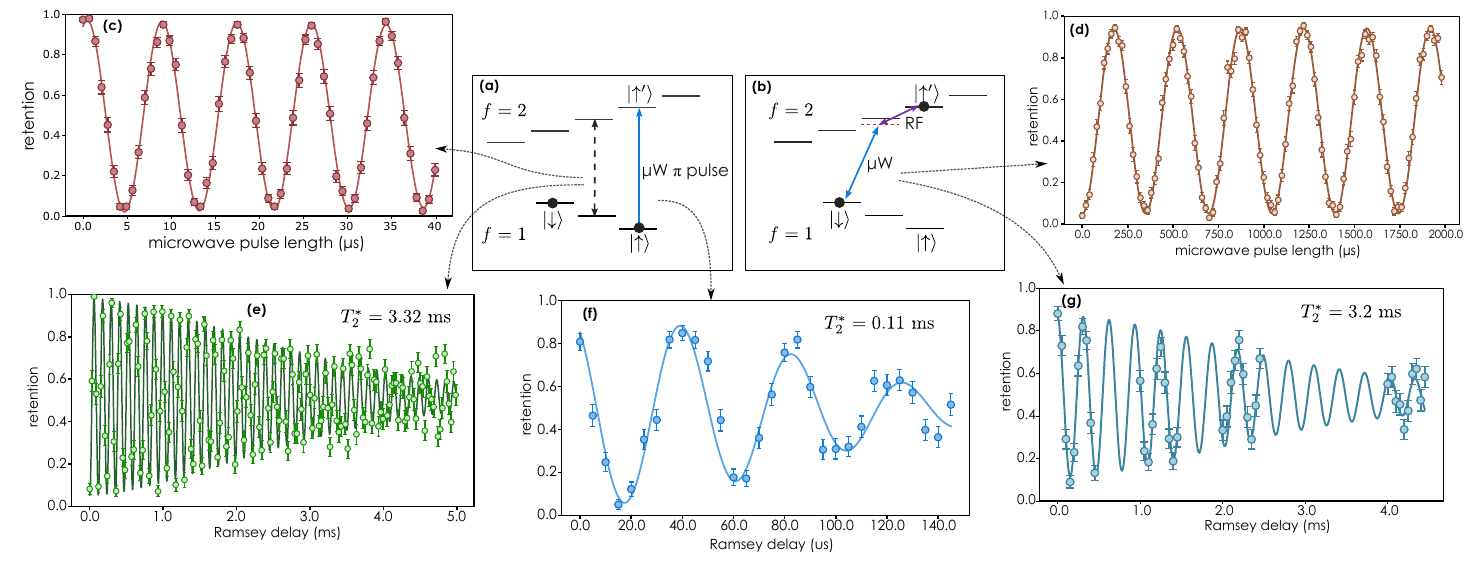}
\caption{(a) After detection of a photon, a microwave $\pi$ pulse of length $5.3~\mu\mathrm{s}$ transfers the population: $\ket{\uparrow} \rightarrow \ket{\uparrow'}$. The new qubit basis $(\ket{\downarrow}, \ket{\uparrow'})$ has a long coherence time at the magic bias field of $\sim3.23~\mathrm{G}$. (b) a two-photon microwave-RF pulse is used to rotate the atomic basis on the Bloch sphere for entanglement verification. (c) Rabi oscillation of the clock transition at rate $118~\mathrm{kHz}$, indicating optical pumping fidelity larger than $0.987(12)$, after correcting for errors in atom measurement. (d) Rabi oscillation for the $\ket{\downarrow} \leftrightarrow \ket{\uparrow'}$ transition driven with a two-photon microwave-RF pulse. The single-photon detuning from $\ket{f=2,m_f=0}$ is $-250~\mathrm{kHz}$ which results in Rabi oscillation at $2.87~\mathrm{kHz}$. The $\pi/2$ pulse length is $87~\mu \mathrm{s}$. (e) Ramsey data on the clock transition $\ket{f=1,m_f=0} \rightarrow \ket{f=2,m_f=0}$ showing a coherence time of $T_2^*=3.32(9)~\mathrm{ms}$. (f) Ramsey results for the transition $\ket{\uparrow} \rightarrow \ket{\uparrow'}$ with a coherence time of $T_2^* = 0.11(1)~\mathrm{ms}$ obtained from the fit. (g) Ramsey fringes of the two-photon transition, indicating $T_2^* = 3.2~\mathrm{ms}$ at $B=3.23~\mathrm{G}$.}
 \label{fig.figure2}
\end{figure*}

\section{Single photon generation and characterization} 
To generate single photons, we follow the scheme first used in Ref.~\cite{Volz2006}. The atom is initially pumped into $\ket{f=1, m_f=0}$ with $\pi$-polarized $795~\mathrm{nm}$ light using one of the on-chip GRIN tubes. The optical pumping light is resonant with the $\ket{5s_{1/2}, f=1} \rightarrow \ket{5p_{1/2}, f'=1}$ transition. A repump laser resonant with the $\ket{5s_{1/2}, f=2} \rightarrow \ket{5p_{3/2}, f'=2}$ transition, combined with the two off-chip MOT beams, avoids population trapping in the $f=2$ manifold.  We achieve an optical pumping fidelity of $0.987(12)$ with a \rsub{$20~\mu\mathrm{s}$} pumping time, which can be seen from the Rabi oscillation of the clock transition in Fig.~\ref{fig.figure2}(c) driven by an on-resonance microwave pulse. 

The initially prepared atom in $\ket{f=1, m_f=0}$ is excited to $\ket{f'=0, m_f'=0}$ with a $\pi$-polarized pulse (Fig.~\ref{fig.figure1} (d)) delivered through the opposite GRIN tube. The excitation pulse has a Gaussian-like temporal shape with FWHM comparable to the lifetime of the excited state ($26~\mathrm{ns}$). The excited atom has three possible decay channels to the $f=1$ manifold with equal probabilities, emitting a photon at $780~\mathrm{nm}$ with three possible polarizations: $\sigma_+, \sigma_-,$ and $\pi$. The $\pi$-polarized light does not couple into the SM fiber due to destructive interference~\cite{Young2022}. Therefore, the chance of collecting a photon from each excitation attempt is at most $2/3$. The trap light is turned off for $250~\mathrm{ns}$ for excitation and photon collection. The photons are detected within a $200~\mathrm{ns}$ window and time-stamped using ARTIQ~\cite{ARTIQ2016}. Figure~\ref{fig.figure1}(e) shows the photon decay-time histogram. The fitted curve is a convolution of the Gaussian excitation pulse with an exponential function with a decay time of $26.4 \pm 0.3~ \mathrm{ns}$, matching the known lifetime of the excited state.

The excited atom decays to the initial state $\ket{f=1, m_f=0}$ with probability of $1/3$. Thus, to increase the success probability of generating a $\sigma_{\pm}$ photon, we repeat the excitation attempt five times, with $20~\mu\mathrm{s}$ delay, after which we apply recooling and optical pumping. We repeat this excitation cycle until the atom is lost from the trap. We check for atom loss after every \rsub{30} excitation cycles. On average, we can repeat the excitation cycle \rsub{$\approx600$} times (which corresponds to \rsub{$5\times 600$} excitation attempts) before losing the atom. The overall single-photon generation, collection, and detection efficiency is measured to be \rsub{$5 \%$} per excitation cycle. We note that the five excitation attempts effectively increase the probability of emitting a $\sigma_{\pm}$ photon in each excitation cycle from $2/3$ to almost unity.

\begin{table*}[!t]
\footnotesize
\caption{Demonstrations of neutral atom-photon entanglement. Success probability is the probability of detecting a photon for each attempted excitation of the atom. Reported fidelities are measured results without correction for separately characterized error sources.   }
\label{tab.atom_photon_entangle}
\centering
\begin{tabular}{|l| l| l| c| c| c|c| }
\hline
Year-group &  Description & Photon encoding & Matter qubit & 
Success probability & Fidelity \\
\hline
2006-Weinfurter \cite{Volz2006} &  lens, $\rm{NA}=0.38$, $\rm{NA}_{eff}=0.29$  & polarization & $^{87}$Rb   & $0.0005$& $0.87 \pm 0.01$\\
2007-Rempe \cite{Wilk2007b}  &  cavity, $C=1.28$  & polarization  & $^{87}$Rb  & 0.013 & $0.86 \pm 0.004$\\
2020-Weinfurter \cite{vanLeent2020} & lens, $\rm{NA}=0.5$ & polarization & $^{87}$Rb &  $0.0075 $ & $0.897 \pm 0.007^d$ \\
2022-Weinfurter \cite{Zhang2022} & lens $\rm{NA}=0.5$ & polarization & $^{87}$Rb &  Trap 1(2)$^b$: $0.00598(0.00144)$ & $0.952(0.941)\pm 0.007$ \\
2024-Rempe \cite{Hartung2024} &  cavity, $C=1.66$  & polarization  & $^{87}$Rb  & 0.33 & $0.866 \pm 0.005$\\
2025-Covey \cite{LLi2025} &  lens, $\rm{NA}=0.63^a$  &  time bin & $^{171}$Yb  & 0.003 & $0.90 \pm 0.014 $ \\
2026  (this work) &   parabolic mirror, $\rm{NA}=0.61$ & polarization &  $^{87}$Rb &$\rsub{0.029}^c$ & $0.93 \pm 0.05 $ \\
\hline
\end{tabular}\\
a) J. Covey, private communication. 
b) Success probability for Trap 2 is lower due to 50 \% loss from 700 m fiber. 
c) The success probability from each excitation cycle (consisting of five back-to-back excitation attempts) is \rsub{$5\%$}, as reported in the main text. However, for fair comparison with the experiments from the Weinfurter group, which uses the same excitation scheme, here we report  the probability of detecting a photon from the first excitation attempt only.
d) This fidelity is for $5~\mathrm{m}$ fiber length without quantum frequency conversion. 
\end{table*}

\rsub{We measured the transmission of the optical setup after the single-mode fiber (see Fig.~\ref{fig.figure1}(a)) and found a total loss of $14.3\%$, including the waveplates, polarizing beamsplitter, and band-pass filters. With a nominal quantum efficiency of $0.65$ for our SPCMs, and using the measured $14.3\%$ loss, the photon collection and coupling efficiency at the output of the SM fiber is estimated to be $\sim 9\%$. This is reasonably close to the $12\%$ expected theoretically for circularly polarized emission collected with a parabolic mirror of $\mathrm{NA}=0.61$ and coupled into a single-mode fiber~\cite{Young2022}. Thus, this design shows a significant improvement over previously reported non-cavity neutral-atom experiments (see Table~\ref{tab.atom_photon_entangle}). In addition to the data reported in Table \ref{tab.atom_photon_entangle} we measured an atom-photon entanglement rate of $12~\rm s^{-1}$ averaged over many experimental cycles including the operations in Fig. \ref{fig.figure1}(c) such as atom reloading and cooling, without entanglement verification. We have not indicated corresponding entanglement rates for the other demonstrations in Table \ref{tab.atom_photon_entangle} since it was not possible to make direct comparisons given non-uniform reporting of experimental procedures.   }

The purity of the single photons is characterized by $g^{(2)}(0)$ given by
\begin{equation}\label{g2}
g^{(2)}(0) = \frac{P_{12}}{P_1 P_2},
\end{equation}
where $P_i$ is the probability of a detection event from detector $i$, and $P_{12}$ is the probability of a joint detection event. These are measured by a standard Hanbury Brown and Twiss experiment~\cite{HanburyBrown1956}, \rsub{with the setup shown in Fig.~\ref{fig.figure1}(a)} with a 50:50 beamsplitter. We measured $g^{(2)}(0)=0.006 \pm 0.006 \ll 1$ indicating high single photon purity.

\section{Atom-photon entanglement} 
The emitted photons are entangled with a single atom in the form~\cite{Volz2006}
\begin{equation}\label{entangled_state}
\ket{\psi^+}= \frac{  \ket{\downarrow, \sigma_+} + \ket{\uparrow, \sigma_-}} {\sqrt2},
\end{equation}
where $\ket{\uparrow}=\ket{f=1,m_f=1}$ and $\ket{\downarrow}=\ket{f=1,m_f=-1}$. This atomic qubit basis is susceptible to magnetic field fluctuations and is expected to have a coherence time similar to the $(\ket{\uparrow},\ket{\uparrow'})$ qubit, for which we measured a coherence time of $T_2^* = 100$–$150~\mu\mathrm{s}$ (see Appendix \ref{app.errors}). To increase the lifetime of the entangled state, we map $\ket{\uparrow}$ to $\ket{\uparrow'}=\ket{f=2,m_f=1}$ using a microwave pulse of length $5.3~\mu\mathrm{s}$ (Fig.~\ref{fig.figure2} (a)). This pulse is applied with a $5~\mu\mathrm{s}$ delay after registering a photon at one of the SPCMs. The $5~ \mu\mathrm{s}$ delay is due to the limitation of real-time decision branching of ARTIQ and can be improved with modified gateware \cite{Stephenson2020}.

The new atomic basis $(\ket{\downarrow},\ket{\uparrow'})$ is an excellent qubit as it has a long coherence time at a bias field of $B=3.23~\mathrm{G}$~\cite{Harber2002}. At this field, the two hyperfine states experience the same first-order Zeeman shift. As shown in Fig.~\ref{fig.figure2}(g), we measured $T_2^* = 3.2~\mathrm{ms}$ for this qubit, without any attempts to control or cancel environmental field fluctuations. This coherence time is expected to be primarily limited by motional decoherence. A long coherence time of $T_2^* > 1~\mathrm{s}$ has been demonstrated for this qubit at lower temperatures in a Bose-Einstein condensate~\cite{Treutlein2004}.

Given the joint atom–photon density matrix $\rho$, the fidelity of the atom–photon state with respect to the Bell state $\ket{\psi^+}$ is
\begin{equation}\label{fidelity_rho}
F = \frac{\rho_{\uparrow'H,\uparrow'H} + \rho_{\downarrow V,\downarrow V}}{2} + \rm{Re}(\rho_{\uparrow'H, \downarrow V}).
\end{equation}
All elements of the density matrix $\rho$ can be obtained from full quantum state tomography of the atom-photon state. Alternatively, one can obtain a lower bound to the entanglement fidelity using only the diagonal elements of the density matrix in the \(z\)- and \(x\)-bases, represented by $\rho$ and $\tilde{\rho}$, respectively~\cite{Blinov2004}:
\begin{equation}\label{fidelity_low}
\begin{split}
F_{\mathrm{low}} &= \tfrac12\Big(
  \rho_{\uparrow' H,\uparrow' H} + \rho_{\downarrow V,\downarrow V}
  - 2\sqrt{\rho_{\uparrow' V,\uparrow' V} \, \rho_{\downarrow H,\downarrow H}}
\\
&\quad
  + \tilde{\rho}_{\uparrow' H,\uparrow' H}
  + \tilde{\rho}_{\downarrow V,\downarrow V}
  - \tilde{\rho}_{\uparrow' V,\uparrow' V}
  - \tilde{\rho}_{\downarrow H,\downarrow H}
\Big).
\end{split}
\end{equation}
The correlation populations $P(\uparrow',H) = \rho_{\uparrow'H,\uparrow'H}$ and $P(\downarrow, V) =  \rho_{\downarrow V,\downarrow V}$ are measured as the correlation between the polarization of the emitted photons and the populations of the atoms in the $(\ket{\downarrow},\ket{\uparrow'})$ basis. This constitutes the \(z\)-basis measurement. The atomic state-selective detection is performed by blowing away atoms in the ground-state $f=2$ manifold with one of the external MOT beams resonant to the $f=2 \rightarrow f'=3$ cycling transition. Thus, a subsequent fluorescence readout determines the state of the atom.

\begin{figure}[!t]
\includegraphics[width=\columnwidth]{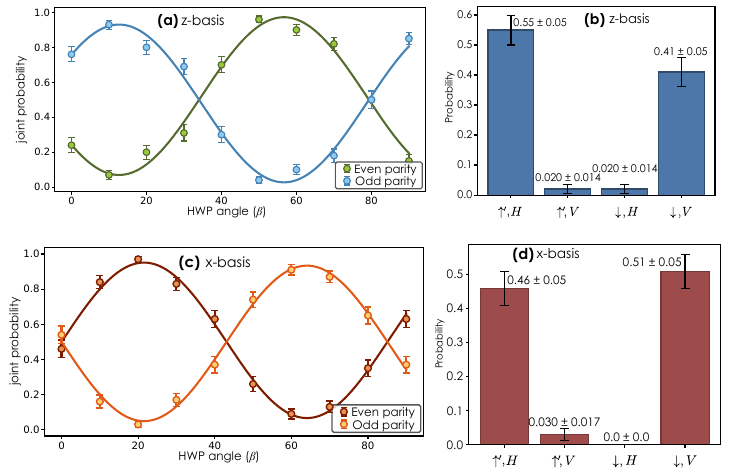}
 \caption{(a) and (c) parity oscillations in \(z\)- and \(x\)-basis, respectively. Even parity: $P(\uparrow',H) + P(\downarrow,V)$, odd parity: $P(\uparrow',V) + P(\downarrow,H)$. The solid curves are fit to sinusoidal functions. (b) and (d) measured joint probabilities in \(z\)- and \(x\)-basis, respectively, at HWP angles giving the maximum visibility in the corresponding parity oscillation.}
 \label{fig.figure3}
\end{figure}

Polarization measurement on the photon is performed using the waveplates and a PBS, as shown in Fig.~\ref{fig.figure1} (a), which projects the polarization onto the H/V basis. Figure~\ref{fig.figure3}(a) plots the joint probability $P(\text{atom},\nu)$ as a function of the HWP angle $\beta$ in the \(z\)-basis, where the atom state $ \in \{ \downarrow, \uparrow'\}$, and the photon state $\nu \in \{ H,V\}$. The SM fiber applies an unknown unitary operation on the polarization of the photons. We do not measure or compensate for the effect of the SM fiber. Instead, we find a QWP angle $\alpha$ that maximizes the amplitude of the parity oscillations in Fig.~\ref{fig.figure3}. 

To measure the remaining density matrix elements required for Eq.~(\ref{fidelity_low}), we use the waveplates to measure the polarization of the photons in the  \(x\)-basis. Similar to the \(z\)-basis measurements, we scan the waveplate angles to maximize the parity oscillations. 

The atomic basis is rotated to the \(x\)-basis using a two-photon microwave-RF $\pi/2$ pulse resonant with the $\ket{\uparrow'} \rightarrow \ket{\downarrow}$ transition. The single-photon transition detuning is $-250~\mathrm{kHz}$, below the $\ket{f=2,m_f=0}$ state. We measure a two-photon Rabi frequency of $2.87~\mathrm{kHz}$ (Fig.~\ref{fig.figure2}(d)), which results in a $\frac{\pi}{2}$-pulse time of $87~\mu \mathrm{s}$. After the $\pi/2$ rotation, the atomic state is measured by blowing away the $f=2$ manifold as described earlier. The correlations between photon polarization and atom state in the \(x\)-basis are shown in Fig.~\ref{fig.figure3}(c) and (d), which yield an entanglement fidelity $F>F_{\mathrm{low}} = 0.93 \pm 0.05$.  

\begin{table}[!t]
\footnotesize
\caption{Main sources of atom-photon entanglement infidelity with their corresponding estimated contributions.}
\label{tab.errors}
\centering
\begin{tabular}{|l| c| }
\hline
source of error &  contribution  \\
\hline
atomic state measurement &  $0.05 \pm 0.07$ \\
atom basis rotation &  $(5\pm5)\times 10^{-3}$ \\
atom dephasing &  $<5\times 10^{-3}$ \\
photon detection noise &  $<3\times10^{-3}$ \\
waveplate rotation error &  $<3\times10^{-3}$ \\
excitation polarization &  $<5\times10^{-3}$ \\
imperfect optical pumping &  negligible \\
\hline
\end{tabular}\\
\end{table} 
\normalsize

\section{Sources of errors}
The possible sources of error in the atom-photon entanglement fidelity and their estimated contributions are shown in Table~\ref{tab.errors}. The main error comes from state-selective atom measurement, which consists of a microwave $\pi$ pulse for $\ket{\uparrow} \rightarrow \ket{\uparrow'}$ mapping, a blow-away pulse for atoms in the $f=2$ manifold, and fluorescence readout of the atom remaining in the trap. The blow-away and fluorescence readout are performed with fidelities of $0.992$ and $0.996$, respectively. The fidelity of the microwave mapping pulse is lower at $0.96(7)$ which results in the overall state-selective atom measurement fidelity of $0.95(7)$. \rsub{After correcting for this measurement error, the central value of the inferred fidelity shifts to approximately 0.98. However, the uncertainty of the readout correction is substantial, so this corrected value should be regarded only as an estimate.} See Appendix \ref{app.errors} for more details. The microwave mapping pulse suffers from a larger error (and larger uncertainty in the error) mainly due to the drift of the microwave resonance frequencies which is caused by polarization drifts of the dipole trap light in the SM fiber. Since this error affects all measurements, in both \(z\)- and \(x\)-basis, it contributes directly to atom-photon entanglement infidelity. 

The coherence time of the initial atomic qubit is relatively short (approximately $110~\mu\mathrm{s}$) before mapping to the new basis $(\ket{\downarrow},\ket{\uparrow'})$. Therefore, the atomic qubit starts dephasing quickly from the photon detection time until we map the atom to the new basis. The microwave mapping pulse is $5.3~\mu\mathrm{s}$ long, and is applied with a $5~\mu\mathrm{s}$ delay after photon detection, limited by the ARTIQ gateware architecture. Therefore, the atomic qubit dephases for approximately $7.7~\mu\mathrm{s}$ on average. However, the \(z\)-basis measurements  are not affected by this dephasing. Hence, we estimate that the limited coherence time of the initial atomic qubit lowers the atom-photon entanglement fidelity by $<5\times10^{-3}$ through the \(x\)-basis measurements.

To first order, the frequency of the two-photon transition $\ket{\downarrow} \leftrightarrow \ket{\uparrow'}$ is insensitive to magnetic noise and trap polarization. Therefore, we estimate that the atomic qubit rotation to the \(x\)-basis contributes only $\sim 5\times10^{-3}$ to the infidelity. Other sources of error, such as false photon detections, errors in the waveplate angles, and misalignment between the polarization of the excitation pulse and the quantization axis, also contribute to atom–photon infidelity, as shown in Table~\ref{tab.errors}. These effects are discussed in more detail in Appendix \ref{app.errors}.  

\section{Summary and outlook}
We have realized a cavity-free neutral atom quantum networking node based on a parabolic mirror that simultaneously serves as a high-NA trapping and collection optic. The node combines high single-photon collection efficiency with high atom–photon entanglement fidelity in a compact, fiber-integrated, and mechanically robust architecture built from millimeter-scale in-vacuum optics on a monolithic platform. Although we have implemented two independent nodes with comparable performance in atom loading and single-photon generation, all quantitative results reported in this manuscript are obtained from a single node. \rsub{See Appendix~\ref{app.node2} for  representative data from the second node. }

The relatively large success probability for single-photon generation directly supports high-quality atom–photon entanglement, suppressing the impact of detector dark counts and background scattering and enabling rapid optimization and stabilization of experimental parameters. \rsub{Although cavity-based systems can achieve higher success probabilities, our results show that a cavity-free architecture can nevertheless reach a regime of efficient photon generation and high entanglement fidelity while retaining a simpler and scalable light-matter interface.} Together with the millisecond-scale coherence time of the memory qubit, these features are well matched to the requirements for efficient meter-scale links and provide a realistic starting point for demonstrations of remote entanglement between two such nodes. \rsub{More generally, the parabolic mirror geometry can be engineered with numerical apertures exceeding unity and can even encompass the emitter, as has been demonstrated with trapped ions~\cite{Maiwald2012}. Extending our architecture to such extreme-NA configurations could further increase the photon collection efficiency while retaining the mechanical robustness of the monolithic design, although such implementations would require additional engineering for optical access. At the same time, for the $\sigma$-polarized photons considered here, large collection angles lead to a spatially varying polarization structure of the collected field, which fundamentally limits the achievable coupling into a single-mode fiber~\cite{LLuo2009}.}

Finally, the in-vacuum assembly is explicitly designed to be compatible with additional high-NA objective lenses for generating and imaging atomic arrays at each node. In the envisioned architecture, one site of the array will be dedicated to coupling to photons and establishing inter-node links, while local Rydberg gates and atom-transport operations distribute this entanglement among multiple qubits within the array. This would enable remote entangled registers shared between different nodes and, when combined with telecom interfaces and multiplexed protocols, provide a route toward scalable neutral atom quantum repeaters and distributed quantum information processing based on cavity-free, parabolic-mirror nodes.

\section{acknowledgment} 
We are grateful to Chris Young and Arian Noori for technical contributions in the early stages of the project and Trent Graham for helpful discussions. This material is based upon work supported by NSF Award No. 2016136 for the QLCI center Hybrid Quantum Architectures and Networks, the U.S. Department of Energy Office of Science National Quantum Information Science Research Centers as part of the Q-NEXT center, and NSF Award No. 2228725.

\newpage

\appendix 

  


\section{Details of the error analysis}
\label{app.errors}

\begin{figure}[!t]
\includegraphics[width=\columnwidth]{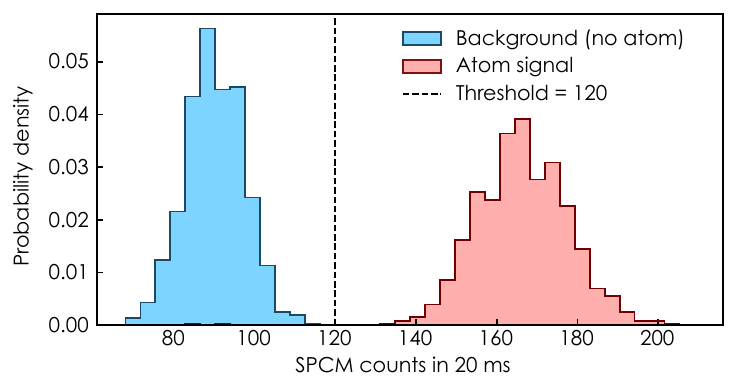}
 \caption{Single-atom fluorescence readout composed of two experiments. The SPCMs counts are obtained by adding the number of counts from the two SPCMs with $20~\mathrm{ms}$ exposure time. In $0.4\%$ of the trials the atom signal lies below the threshold as can be seen from the few red evens underneath the blue histogram. These events are from atom loss during the readout sequences.}
 \label{fig.A_Readout}
\end{figure}

During the atom loading process, we monitor the photon counts on the SPCMs constantly with $20~\mathrm{ms}$ exposure time to look for an atom signal which shows up as a clear increase in the photon counts. Upon measuring photon counts above a threshold, we turn off the MOT beams and the magnetic field for $50~\mathrm{ms}$ to dissipate the MOT while the single atom is kept in the dipole trap. Then, we perform the first fluorescence readout to ensure the trapping of a single atom. With this method, we achieve an atom loading probability of more that $95\%$, i.e. the first readout indicates a single atom in more than $95\%$ of the shots. To measure the fidelity of the atom fluorescence readout we performed two experiments. First, we measured the background signal without loading any atoms in the trap (blue histogram in Fig.~\ref{fig.A_Readout}). Then, in the second experiment, we loaded single atoms and performed a second readout if the first readout indicated an atom in the trap (red histogram in Fig.~\ref{fig.A_Readout}). Ideally, we expect to see the atom in the second readout in all of the measurements. However, in $0.4\%$ of the trials, the signal was lower than the threshold (see the red data points underneath the blue histogram in Fig.~\ref{fig.A_Readout}). This data indicates a fluorescence readout fidelity of 0.996. The atom loss between the two readouts is mainly caused by background pressure which limits the trap lifetime to a couple of seconds.

To measure the fidelity of the blow-away pulse, we performed four experiments in which we pumped the atoms into $f=1$, and $f=2$ manifolds, then measured the retention with a fluorescence readout with and without applying a blow-away pulse. From these experiments, we measured a blow-away fidelity of minimum $0.988(4)$. Combining the blow-away fidelity with the fluorescence readout fidelity, we arrive at hyperfine-selective readout fidelity of $0.988(3)$. 

The main source of error in our state-selective atom readout comes from the microwave mapping pulse to transfer $\ket{\uparrow}$ to $\ket{\uparrow'}$. Since the dipole trap light is delivered to the atom through the SM fiber, it does not have a good linear polarization at the atom. By adjusting the $852~\mathrm{nm}$ waveplates before the SM fiber, and maximizing the transmission through the polarizer and the multi-mode fiber at the back of the parabolic mirror, we optimize the trap polarization to be linear. However, this method is not very accurate and we expect a couple of degrees of ellipticity which can fluctuate by environment temperature and mechanical vibrations. The circular component of the trap light can shift the energy levels of the atomic ground state through the finite vector polarizability of the atoms and act as a fictitious magnetic field~\cite{Thompson2013b}. Hence, fluctuation in the polarization of the trap light results in fluctuation of the microwave transition frequencies, especially between $\ket{\uparrow}$ and  $\ket{\uparrow'}$ as they are both magnetically sensitive. To mitigate this problem, we lower the trap power adiabatically for the microwave transitions to about $100~\mu \mathrm{K}$. See the experiment sequence in Fig.~\ref{fig.figure1}(c) of the main text. In addition, we covered the SM fiber with a thick plastic jacket to protect it from the air flow in the laboratory. By tracking the microwave transition frequency of $\ket{f=1,m_f=0} \rightarrow\ket{f=2,m_f=1}$ transition over several hours, we observed that drift in the transition frequency is often less than $1~\mathrm{kHz}$ which is significantly smaller than the width of the transition, on the order of a few tens of kHz. 

We measured the fidelity of the microwave mapping pulse by first optically pumping atoms into $\ket{f=1,m_f=0}$ and mapping them to $\ket{f=2,m_f=1}$ by a microwave $\pi$ pulse. We then measured atom retention after a blow-away pulse for both on-resonance and far off-resonance microwave pulses. Correcting the results for the errors in optical pumping, blow-away, and atom readout, we estimate the fidelity of this microwave mapping pulse to be $0.98(7)$. In microwave mapping from $\ket{f=1,m_f=1}$ to $\ket{f=2,m_f=1}$, however, both states are magnetically sensitive. Therefore, we assume that the infidelity of this mapping is twice the infidelity of the $\ket{f=1,m_f=0} \rightarrow \ket{f=2,m_f=1}$ mapping. Hence, the fidelity of the overall state-selective atom measurement, which consists of microwave mapping from $\ket{f=1,m_f=1}$ to $\ket{f=2,m_f=1}$, blow-away pulse, and fluorescence readout, is estimated to be $0.95(7)$.

Error in microwave mapping pulses can also affect the basis rotation on the Bloch sphere to \(x\)-basis which is performed by a two-photon microwave-RF pulse. However, the resonant frequency for $\ket{\uparrow} \rightarrow \ket{\uparrow'}$ transition is insensitive, to the first order, to magnetic noises as well as to the trap polarization errors. We measured the performance of the two-photon transition by optically pumping atoms in $\ket{f=1,m_f=0}$ state and transferring them to $\ket{\uparrow'}$ using a microwave $\pi$ pulse before transferring them to $\ket{\downarrow}$ with the two-photon pulse. Then, we performed hyperfine-selective readout and measured that the population transfer from $\ket{\uparrow'}$ to $\ket{\downarrow}$ can be performed with $0.96(3)$ fidelity. We assume that the population transfer infidelity in the $\pi$ pulse is due to the pulse area error and is twice the corresponding error for the $\pi/2$ pulse. This error affects the measurements in \(x\)-basis only. By propagating this error in the area of the $\pi/2$ pulse to the atom-photon entanglement fidelity calculation, we arrive at $(5\pm0.5)\times10^{-3}$ for the contribution of imperfect atom basis rotation. 

Detection of the photon projects the atom into a superposition in $(\ket{\downarrow}, \ket{\uparrow})$ qubit basis which are both magnetically sensitive. Therefore, the atomic qubit starts dephasing rapidly until it is mapped to the magic qubit basis $(\ket{\downarrow}, \ket{\uparrow'})$. We time-stamp the photon arrival time and apply the microwave mapping pulse with exactly a $5~\mu\mathrm{s}$ delay (within $\approx 1~\mathrm{ns}$ precision of the electronics) after the photon arrival time. The $5~\mu\mathrm{s}$ delay time is required in ARTIQ to time-stamp the photon and trigger the mapping pulse, but this delay can be improved using a customized gateware~\cite{Stephenson2020}. The mapping pulse is $5.3~\mu\mathrm{s}$ long. Therefore, we assume that the atomic qubit dephases for about $7.7~\mu\mathrm{s}$ on-average. In the experiment, we do not drive the transition $\ket{\downarrow} \rightarrow \ket{\uparrow}$. Hence, we measured the coherence time of $\ket{\uparrow} \rightarrow \ket{\uparrow}$ transition using a Ramsey sequence which is expected to have a similar coherence time. Figure~\ref{fig.figure2}(f) of the main text shows the result of the Ramsey experiment with a fit function $\propto e^{-(t/T_2^*)^2} \cos{(\omega t + \phi)}$ and $T_2^*=0.11(1)~\mathrm{ms}$. Therefore, after $7.7~\mu\mathrm{s}$, the atomic coherence reduces by $<1~\%$. Since the dephasing affects the \(x\)-basis measurements only, it adds an error $<5\times10^{-3}$ to the atom-photon entanglement fidelity. 

If the excitation pulse is too short (broad in spectrum) it can off resonantly couple to higher excited states and lower the entanglement fidelity. The next excited state is $\ket{f'=1}$ which is only 72 MHz away from $\ket{f'=0}$. However, the transition $\ket{f=1,m_f=0} \rightarrow \ket{f'=1,m_f'=0}$ is forbidden. The 2nd possible excited state ($\ket{f'=2}$) is 229 MHz away. Considering the $18~\mathrm{ns}$ width of our excitation pulse, the probability of off-resonant excitation to $\ket{f'=2}$ level is negligibly small.

The excitation and atom-photon entanglement scheme are insensitive to small errors in the optical pumping fidelity. This is because the excitation pulse does not couple well to $\ket{f=1,m_f=\pm 1}$ states. However, if the polarization of the excitation pulse is not well aligned to the quantization axis, the atom can off-resonantly excite to $\ket{f'=1}$ manifold either from the initial state $\ket{f=1,m_f=0}$ to $\ket{f'=1,m_f'=\pm 1}$ or from $\ket{f=1,m_f=\pm 1}$ states which can be occupied after the first excitation pulse. In this scenario we expect to detect multiple photons within the 5 excitation attempts. However, only in $0.5\%$ of the excitation cycles, more than one photon is experimentally detected within the 5 excitation attempts. Thus, we can safely assume that the error in the polarization of the excitation pulse does not contribute more than $0.5\%$ to atom-photon entanglement infidelity.

The basis rotation and projective measurements on the photons are performed with a pair of QWP and HWP. The QWP transforms elliptical polarization at the output of the SM fiber to linear polarization. Since the SM fiber applies an unknown unitary operation on the polarization, the QWP angle for the rotated basis is not necessarily at $45^\circ$ from the \(z\)-basis angle. Indeed, the QWP angles for the \(z\)- and \(x\)-basis measurements, reported in Fig.~\ref{fig.figure3} of the main text, are $60^{\circ}$ and $25^{\circ}$, respectively. The visibility of the parity oscillations can reduce if the QWP is not at the optimum angle. Considering the worst case scenario when the polarization at the SM fiber output is circular, deviation of $\Delta\alpha$ from the optimum angle reduces the visibility of the parity oscillations by $\cos(2\Delta\alpha)$. We expect an error of $\Delta\alpha\approx2^\circ$ in the QWP angle which can result in reduction of visibility by $\approx 0.0024$, in the worst case scenario. Similar error is expected for the HWP angle, in both \(z\)- and \(x\)-basis. Propagating these errors to the fidelity calculation, we conclude that errors in waveplates angle can contribute up to 0.003 in entanglement infidelity.

Finally, we consider the error due to false counts on the SPCMs. Experimentally, we measured $50$ dark counts per second per SPCM. Since the overall single-photon collection and detection efficiency of our setup is $5\%$, we need to repeat the excitation cycle $100/5$ times to detect a single photon. Considering the gate time of $200~\mathrm{ns}$ and 5 excitation attempts, we expect about $0.3\%$ of the total photon counts to be false. False photon detections trigger atom measurements and affect all the results. However, we may still find the atomic state in correlation with the photon noise counts accidentally. Therefore, the effect of false counts on entanglement fidelity is estimated to be less than $0.003$.

\section{Vacuum system design}
\label{app.vacuum}

\begin{figure}[!t]
\includegraphics[width=\columnwidth]{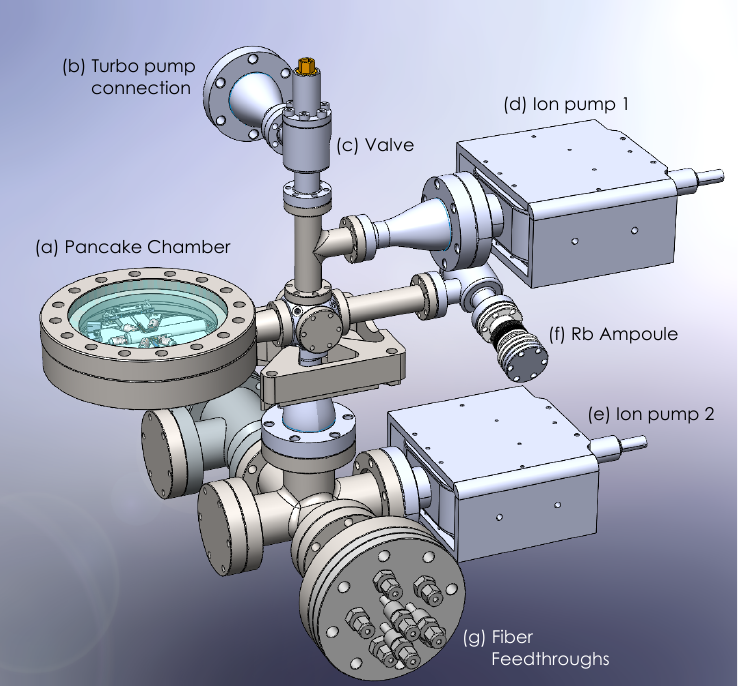}
 \caption{Design of the full vacuum system  (a) Pancake chamber with the optical chip mounted inside. (b) and (c) Valve for turbo pump connection. (d) and (e) Ion pumps for pumping of both upper and lower chamber. (f) Rb ampoule sitting inside the bellows. (g) eight fiber feedthroughs. }
 \label{fig.appendix_figure4}
\end{figure}

The vacuum system consists of the pancake chamber, two ion pumps, a rubidium ampoule, eight fiber feedthroughs, and a valve for connecting a turbo pump (See Fig.~\ref{fig.appendix_figure4}).

\begin{figure*}
\includegraphics[width=\textwidth]{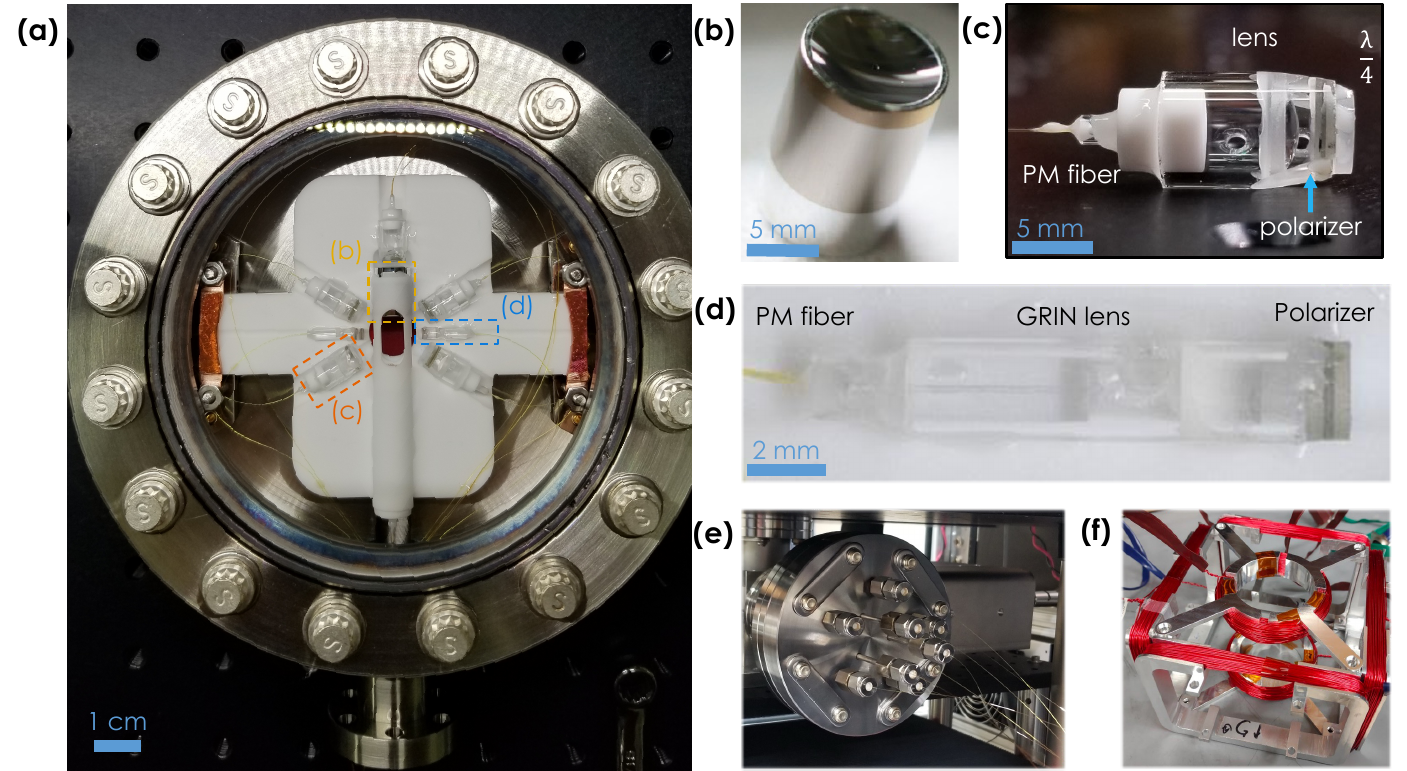}
 \caption{Pancake chamber and on-chip modules (a) Top view of the pancake chamber showing the optical modules mounted on the Macor chip. (b) Parabolic mirror. (c) MOT beam module. (d) GRIN module for optical pumping and excitation. (e) Eight fiber feedthroughs. (f) Coil assembly designed to mount around the pancake chamber.}
 \label{fig.appendix_figure1}
\end{figure*}

The pancake chamber is custom designed to securely mount the optical chip. We refer to it as the “pancake chamber” due to its low-profile, disk-like geometry (See Fig.~\ref{fig.appendix_figure1}(a) for a close-up view of the pancake chamber.). The chip is mounted using two groove grabbers and oriented to minimize mechanical stress on the SM fiber that delivers the dipole-trap beam. The chamber has fused silica viewports on both sides to provide external optical access, each anti-reflection coated over $400$--$1100~\mathrm{nm}$. The top viewport has a larger window (3.8" diameter), while the bottom viewport has a smaller window (1.4" diameter). The viewport thickness ($7.36~\mathrm{mm}$) was chosen to compensate for optical aberrations when used with existing custom high-NA objective lenses (JenOptik), enabling future integration of an optical tweezer array.

A total of eight optical fibers are routed from the pancake chamber - the upper part of the vacuum assembly down through the chamber to the fiber feedthroughs at the bottom, where they exit the vacuum system (Fig.~\ref{fig.appendix_figure4}(g)). At the feedthroughs, the vacuum is sealed by passing each fiber through a Teflon plug inside a Swagelok fitting, which is carefully tightened so that the fiber stays undamaged. During and after bakeout, the fittings had to be re-tightened to achieve a good vacuum pressure to compensate for thermal expansion and contraction of the Teflon plugs as the temperature changed.

To protect the science region from potential leaks near the fiber feedthroughs, the upper and lower part of the vacuum system are separated by a differential pumping region with each part equipped with its own ion pump (See Fig.~\ref{fig.appendix_figure4}(d) and (e)). In this region, a copper pinch-off tube - $7~\mathrm{cm}$ in length with an inner diameter $6.35~\mathrm{mm}$ - was used in place of a standard conflat copper gasket. The middle section of the tube was pinched down to a slit of approximately $1~\mathrm{mm}$ wide, reducing the conductivity between two regions to $0.44~\mathrm{L/s}$, which is well below the upper ion pump’s pumping speed of $10~\mathrm{L/s}$. The conductivity between the science region and the lower chamber is intrinsically lower due to the right angle geometry between the science chamber and differential pumping path. As a result, we achieve excellent vacuum levels in both regions, $P_{\mathrm{upper}} = 4\times 10^{-10}~\mathrm{Torr}$, $P_{\mathrm{lower}} = 2 \times 10^{-9}~\mathrm{Torr}$, with the upper region closer to the science region maintaining the lower pressure.

\section{On-chip architecture}
\label{app.chip}

We present a compact and plug-and-play node design in which most of the optics are pre-aligned in vacuum on a chip with fiber interfaces. The in-vacuum optics are mounted on a custom chip composed of a Macor substrate, which was chosen for its low thermal expansion, mechanical strength, and machinability. The Macor chip is fabricated with precision-milled slots to accommodate different optical modules, including four out of six MOT modules for laser cooling, a dipole trap module for both atom trapping and photon collection, two Gradient-index (GRIN) lens modules for optical pumping and excitation, and a multimode (MM) fiber module for checking the polarization of the incident dipole trap beam. All modules are secured on the chip using a low-outgassing, UHV-compatible epoxy. This modular pre-aligned design reduces post-alignment overhead, requiring only two external MOT beams outside the chamber to operate the device.

The chip incorporates several types of optical modules, each individually assembled and aligned to perform a specific function (See Fig.~\ref{fig.appendix_figure1}). All modules share a common structure, consisting of a polyimide-coated fiber inserted through a fiber ferrule and a collimating lens, but differ in size and include additional optical elements tailored to their respective roles. The optical components are mounted inside glass or Macor tubes that include venting holes to prevent air from becoming trapped inside during vacuum bakeout.

\subsection{MOT beam module}
The MOT beam module delivers the cooling and repump beams required to form a MOT (See Fig.~\ref{fig.appendix_figure1}(c)). Each module includes a PM fiber inserted through a fiber ferrule, a collimating lens ($f = 5~\mathrm{mm}$), a polarizer aligned to the slow-axis of the fiber, and a quarter waveplate to generate the required circular polarization. Four MOT beam modules are integrated on the chip, and together with two external MOT beams, they form a standard six-beam MOT configuration used for cooling and repump. 

\subsection{GRIN modules}
The GRIN modules are used for optical pumping and excitation of the atom (See Fig.~\ref{fig.appendix_figure1}(d)). Each module incorporates a PM fiber inserted through a fiber ferrule, a GRIN lens for collimation (effective focal length of $4~\mathrm{mm}$) and a polarizer aligned to the quantization axis.

\subsection{Parabolic mirror module}
The parabolic mirror module consists of an SM fiber inserted through a fiber ferrule, a collimating lens ($f = 54~\mathrm{mm}$), and a high-NA (0.61) parabolic mirror with a diameter of $9~\mathrm{mm}$ and a working distance of $5.26~\mathrm{mm}$, \rsub{defined here as the axial distance from the center of the mirror to the focal point.} The parabolic mirror is made of aluminum with silver coating (See Fig.~\ref{fig.appendix_figure1}(b)). This module delivers the dipole trap beam and simultaneously collects photons emitted by the atom along the same optical axis. We use an SM fiber rather than a PM fiber to avoid introducing significant birefringence onto the collected photons, which would otherwise destroy the atomic-photon entanglement. As a result, however, the polarization of the dipole trap light is not maintained through the SM fiber.

\subsection{Multi-mode fiber module}
The multi-mode fiber module enables polarization monitoring of the dipole trap beam. It consists of a multi-mode fiber inserted through a fiber ferrule, a collimating lens ($f = 5~\mathrm{mm}$), and a polarizer. The module is positioned behind the parabolic mirror to collect a small amount of dipole trap light that passes through the mirror’s $0.5~\mathrm{mm}$ on-axis through-hole and the polarizer. Because multi-mode fibers efficiently gather this transmitted light, independent of mode quality, this module provides a signal for checking and aligning the dipole trap beam polarization, compensating for the lack of polarization preservation in the SM fiber used in the parabolic mirror module. However, we use this signal to align the polarization of the dipole trap light only when the environmental temperature changes significantly. 

\subsection{Beam module to fiber interface}
To facilitate pre-alignment and on-chip assembly, the fibers attached to each optical module are initially connectorized. After alignment on the chip, the connectors are removed and the fibers are cut to the appropriate length to pass through the vacuum system—from the chip down to the fiber feedthroughs (Fig.~\ref{fig.appendix_figure1} (e)). After exiting the feedthrough, each fiber is fusion-spliced (Fujikura FSM-100P) to either a SM or a PM fiber, depending on the optical requirements of the corresponding module.

\subsection{Prebake of beam modules}
Although the epoxy is UV-curable, UV exposure alone does not fully harden it, so a thermal bake is required to complete the cure and ensure long-term mechanical stability before on-chip assembly. Once all modules were assembled, they were baked in air at $120^\circ ~\mathrm{C}$ using a small oven made from heat tapes, with the temperature ramp limited to $0.5^\circ ~\mathrm{C}/\mathrm{min}$ to avoid mechanical stress and preserve the alignment.

\section{Parabolic mirror characterization and alignment}
\label{app.parabolic}

\begin{figure}[!t]
\includegraphics[width=\columnwidth]{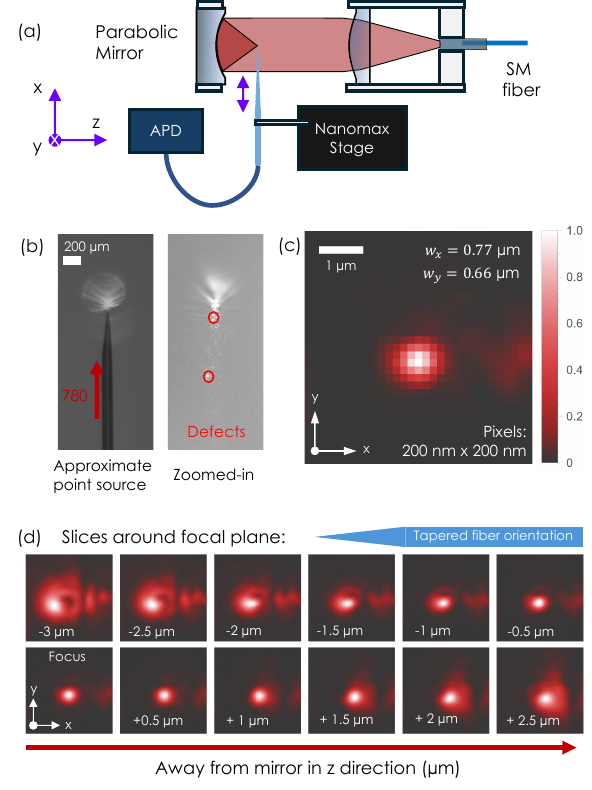}
 \caption{Parabolic mirror characterization using a tapered fiber mounted on a 6-axis stage (Thorlabs Nanomax) (a) Schematic of the setup. (b) Emission pattern of the tapered fiber with coupled $780~\mathrm{nm}$ light (right: longer exposure to reveal light leaking from fiber defects) (c) 2D raster scan at the parabolic mirror focus. (d) 2D raster scans taken at multiple axial positions around the focus. The features appearing around the focal point are artifact of the defects on the tapered fiber as seen in part (b).}
 \label{fig.appendix_figure2}
\end{figure}

One challenge in using the parabolic mirror is characterizing its focal spot as it cannot be measured by standard techniques without obstruction of the impinging light. To address this challenge, we used an near-field optical microscopy (SNOM) uncoated tapered fiber as a point detector to probe the focal spot. Since the bare fiber has a diameter of $\sim 125~\mu\mathrm{m}$, it does not block or disturb the impinging light significantly (Fig.~\ref{fig.appendix_figure2}(a)) and provides a precision of $\sim100~\mathrm{nm}$.

Figure~\ref{fig.appendix_figure2}(b) shows the emission pattern of the tapered fiber when we send $780~\mathrm{nm}$ light through the other end of the fiber. Defects on the tapered fiber are visible when the image is taken with longer exposure time. This indicates that light can also couple into these imperfections when the fiber is used as a point detector. However, coupling into such defects is strongly suppressed and negligible, except when the focal spot is exactly aligned with a defect spot.

Because the focal spot of the parabolic mirror was expected to be less than one micron, sub-micron precision was required for accurate characterization. Thus, we mounted the tapered fiber tip on a 6-axis piezo stage and ran two-dimensional raster scans near the focal region. The output of the fiber was attached to an avalanche photodetector (Thorlabs APD430A) for sensitive measurement of the collected light. The setup is shown in Fig.~\ref{fig.appendix_figure2}(a).

We measured the beam waist at the focus of the parabolic mirror to be $w_x = 0.77~\mu\mathrm{m}$ and $w_y = 0.66~\mu\mathrm{m}$, as shown in Fig.~\ref{fig.appendix_figure2}(c). A full reconstruction of the three-dimensional intensity profile of the focused beam was also obtained from a series of transverse scans, as shown in Fig.~\ref{fig.appendix_figure2}(d). The beam profile appears to be highly aberrated, especially along the x-axis and at planes farther away from the focus. However, this effect is an artifact arising from weak coupling of light into defects on the stem of the fiber tip, visible in Fig.~\ref{fig.appendix_figure2}(b). This was confirmed by rotating the mirror tube by $90^{\circ}$ with respect to the fiber tip and repeating the scan. Subsequently, the aberration reappeared in the same position relative to the fiber tip, confirming that it originates from the fiber itself rather than from the focal intensity distribution.

Precise alignment of the parabolic mirror is essential for achieving diffraction-limited focusing and high collection efficiency. However, the mirror’s alignment is highly sensitive to small angular errors. According to our Zemax simulations, an axial tilt as small as $0.05^\circ$ introduces a noticeable aberration. 

To address this challenge, we employed an intermediate Macor tube that serves as a mechanical interface between the Macor chip base and the small parabolic mirror (The tube is shown in  Fig.~\ref{fig.appendix_figure1}(a) at the center of the chip). The parabolic mirror has a total length of $12~\mathrm{mm}$, with $10~\mathrm{mm}$ seated inside the Macor tube, ensuring a mechanical stability. The mechanical tolerance between the mirror and the tube was set to less than $40~\mu\mathrm{m}$, limited by the different thermal expansion rate between the parabolic mirror (made of aluminum) and Macor.

The Macor tube assembly also houses the achromatic lens and the single-mode fiber which are positioned in the tube before installing the parabolic mirror. The single-mode fiber is first inserted into a glass ferrule, which is then mounted in an adapter (made of Macor) sitting at the end of the tube. The fiber is anti-reflection coated covering both $780~\mathrm{nm}$ and $852~\mathrm{nm}$. An achromatic doublet is installed inside the tube from the opposite side and positioned in front of the fiber to collimate the outgoing beam before it reaches the parabolic mirror. 

\section{On-chip alignment}
\label{app.alignment}

Precise alignment of all optical modules was critical, as the performance of the chip architecture relies on their spatial overlap at the trap position. Once the modules were assembled individually, they were aligned one by one on the Macor chip, which had precision-milled slots serving as mechanical guides and greatly simplifying the alignment process. These slots provided a mechanical tolerance of approximately $25~\mu\mathrm{m}$, meaning that the MOT and GRIN beams were already close to their correct positions once seated properly. The parabolic mirror module, however, required additional care due to its high numerical aperture and correspondingly tighter alignment tolerances.

We began by aligning the MOT beam modules. Each pair of opposing MOT beams was adjusted to maximize cross-fiber coupling, achieving up to $\sim 30\%$ coupling efficiency. Once the MOT beams were aligned, the parabolic mirror module was positioned such that its focal point coincided with the MOT beam overlap. To verify this overlap, we used an acetone-washed transparency sheet ($\sim 80~\mu\mathrm{m}$ thickness as a scattering screen, placed at the focus of the parabolic mirror. The scattered lights from different beams on this screen are monitored by a CMOS camera. For more details, see Ref.~\cite{PHuft2025thesis}. The transparency sheet, mounted on a three-axis translation stage, was inserted from the bottom of the Macor chip with its plane facing the GRIN tubes; in this configuration, it did not obstruct the light incident onto the parabolic mirror.

After aligning the parabolic mirror module, the GRIN tubes were adjusted so that their beams also overlapped with the mirror’s focal point. In addition, the polarization of each GRIN beam was aligned to the quantization axis, which lies in the plane of the Macor chip. Once each module was precisely aligned, we applied a UV-curable, UHV-compatible epoxy and set it with UV light. 

After the full alignment procedure was completed, the entire chip was covered and baked in air at $100^\circ ~\mathrm{C}$ for a few hour, safter which we checked the alignment and did not observe any drift after baking. These checks confirmed that the modules and their alignments survived the pre-bake process without drift.

\section{Node 2 based on parabolic mirror}
\label{app.node2}

\begin{figure}[!t]
\includegraphics[width=\columnwidth]{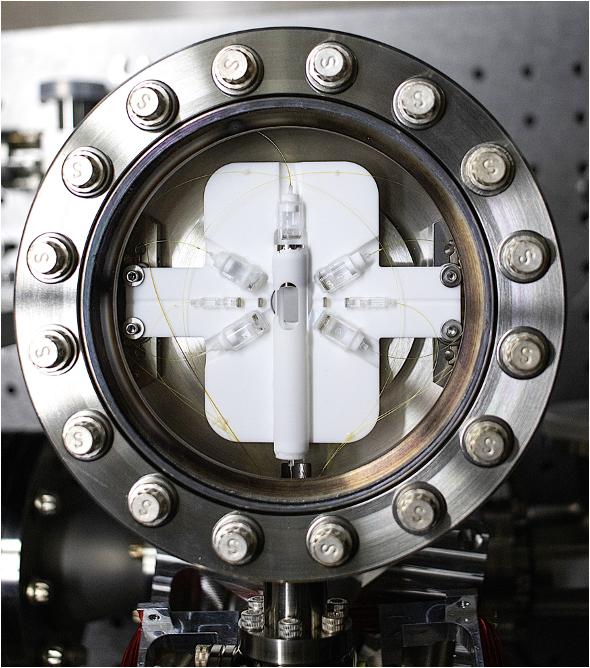}
 \caption{\rsub{Top view of the pancake chamber of Node 2. Several minor differences from Node 1 are visible. The chip is mounted to the groove grabber using only screws and spacers of a specific height, without any additional components. In addition, a small crack on the left side of the parabolic-mirror tube provides a clearer side view toward the center of the chip.}}
 \label{fig.appendix_figure_node2_1}
\end{figure}

\rsub{We built a duplicate of the parabolic-mirror-based node, which we refer to as Node 2 (See Fig.~\ref{fig.appendix_figure_node2_1}). The overall design, including the vacuum system and optical modules, is the same as that of Node 1. There are minor differences arising from mechanical tolerances and alignment quality, but these are not expected to significantly affect its performance. The focal spot of the parabolic mirror was characterized using the same method described in Appendix ~\ref{app.parabolic}. We measured $w_x = 0.75~\mu\mathrm{m}$ and $w_y = 0.7~\mu\mathrm{m}$, which are comparable to the values obtained for Node 1.}

\begin{figure}[!t]
\includegraphics[width=\columnwidth]{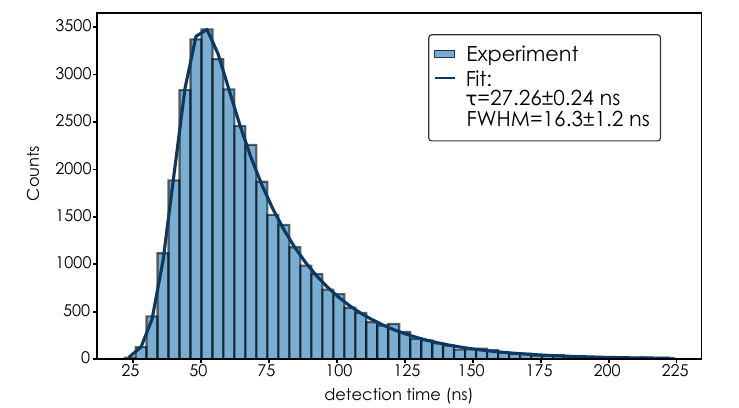}
 \caption{\rsub{Histogram of the single-photon detection time from Node 2. The fit is a convolution of a Gaussian with FWHM of $16.3~\mathrm{ns}$ with an exponential with a decay time of $27.26~\mathrm{ns}$ which agrees to within a few percent with  the known  lifetime of the excited state.}}
 \label{fig.appendix_figure_node2_2}
\end{figure}

\rsub{
Consistent with the nearly identical design, we observe comparable performance in Node 2. We measured $g^{(2)}(0)=0.006 \pm 0.0043 \ll 1$, confirming strong single-photon characteristic. The histogram of the single-photon detection times from this measurement is shown in Fig. ~\ref{fig.appendix_figure_node2_2}.}

\rsub{ 
The experimental sequence is very similar to that of Node 1, with a few key differences. First, optical pumping was performed for
$100~\mu\mathrm{s}$. In addition, atom loss was checked every $30$ excitation cycles, and on average we were able to repeat the cycle 
$256$ times before losing the atom. In this measurement, we used four excitation attempts per cycle. Under these conditions, the single-photon generation rate was $3.65\mathrm{/s}$, including atom loading time. The overall single-photon generation, collection, and detection efficiency was measured to be $3.67\%$ per excitation cycle.}

\rsub{ 
These values have not yet been optimized. However, since Node 2 is an essentially identical duplicate of Node 1, and given the results obtained so far, we expect similar performance to be achievable in Node 2. Measurements to verify atom–photon entanglement in Node 2 are currently in progress.}

\section{MOT switchyard and power stabilization}
\label{app.MOToptics}

\begin{figure*}
\includegraphics[width=.95\textwidth]{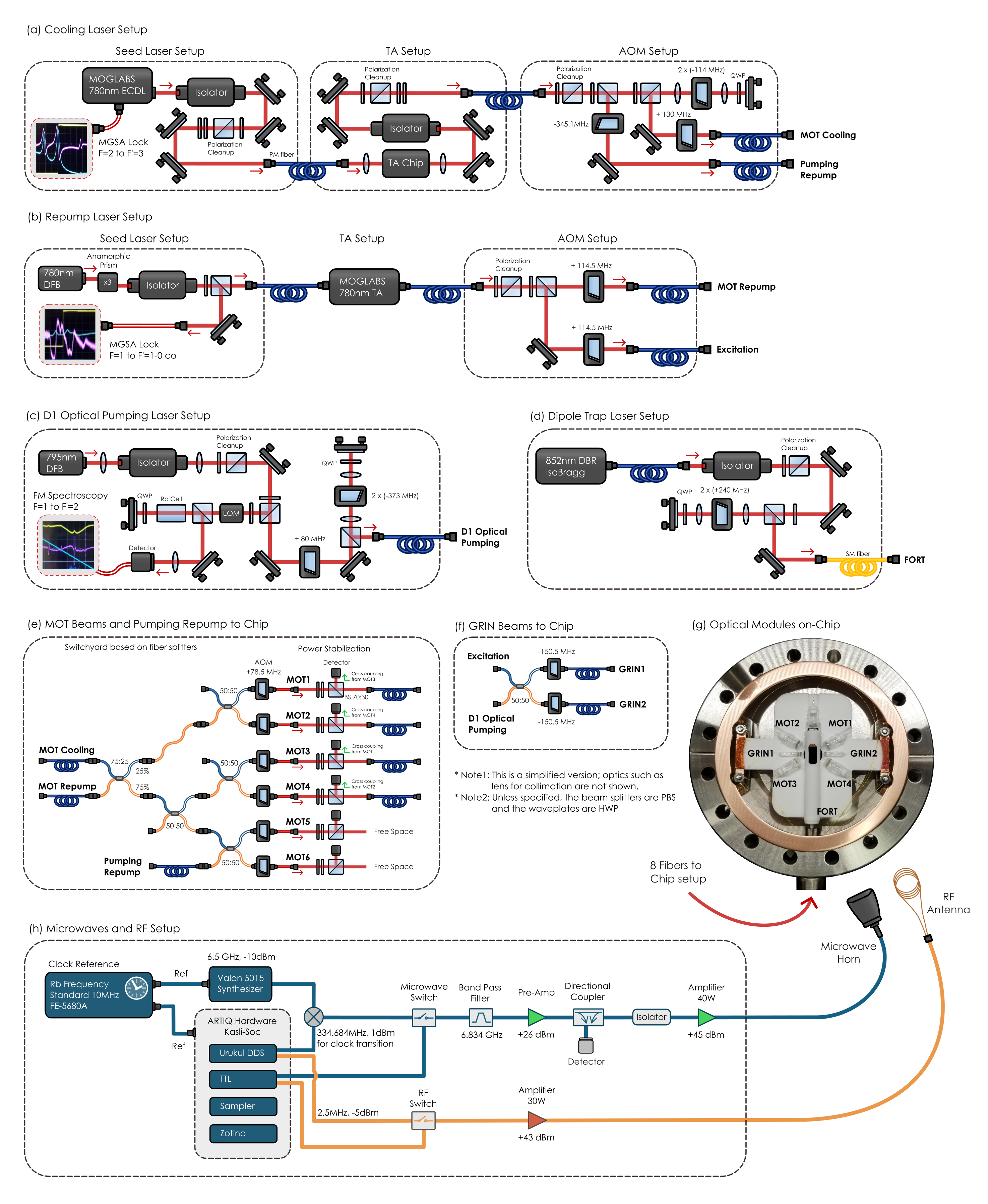}
 \caption{Laser schematic and microwave-RF setup. (a) MOT cooling and pumping-repump setup. (b) MOT repump and excitation setup. (c) D1 optical pumping setup. (d) Dipole-trap laser setup. (e) Fiber switchyard for MOT beams and pumping-repump. (f) GRIN beams to chip - excitation and optical pumping light are combined and routed to both GRIN1/GRIN2. The enabled AOM selects which light is delivered to GRIN1/GRIN2. (g) Optical modules mounted on chip. (f) Microwave and RF setup. Both the microwave horn and the RF antenna are mounted on top of the pancake chamber and they are larger than they appear in this schematic.}
 \label{fig.appendix_figure3}
\end{figure*}

\subsection{Initial beam balancing of MOT beam modules}
Stable MOT formation and efficient polarization-gradient cooling (PGC) require well-balanced beam powers, especially between counter-propagating beam pairs. In our pre-aligned chip architecture, however, the optical powers delivered to the atoms cannot be directly measured. We therefore used atom fluorescence as a proxy for the beam powers and balanced the on-chip beams in situ. 

A CMOS camera equipped with a $780~\mathrm{nm}$ filter was used to image the MOT region. With cooling and repump light applied, each on-chip beam was turned on individually and its power was adjusted by tuning the RF drive to the corresponding fiber AOM until the fluorescence reached a predefined set point. This procedure was repeated for all four on-chip beams to equalize their effective powers at the trap. After the beams were balanced, we measured the corresponding cross-coupled power at the beam sampler and used these values as reference set points for active power stabilization during operation (Fig.~\ref{fig.appendix_figure3}(e)).

\subsection{MOT beam power stabilization}
MOT cooling and repump light are first combined and coupled into a PM fiber, then distributed into six channels using fiber splitters: four for the on-chip MOT modules and two for the external MOT beams (Fig.~\ref{fig.appendix_figure3}(e)). In each channel, a fiber-AOM provides independent power control and fast switching via the applied RF power. Each fiber-AOM is operated at a slightly different RF frequency (by $10~\mathrm{kHz}$) to avoid slowly drifting standing wave and interference between opposite MOT beams. 

After the fiber-AOMs, the light enters a short free-space section containing wave plates for polarization cleanup and a beam sampler. The sampler transmits the forward-going beam, which is coupled into fiber and delivered to the corresponding on-chip module. A small fraction of the light is cross-coupled into the opposing module and returns through the opposing fiber channel. This returning light is picked off by the beam sampler and detected on a photodiode. We use this signal as an in-line power monitor for each channel and actively stabilize the power via feedback to the fiber-AOM RF drive. This power-stabilization routine is applied sequentially to each beam module at the start of the experiment.

\subsection{Dipole-trap power stabilization}

Stable dipole-trap operation requires maintaining constant laser power. To monitor the power of the trap light delivered to the vacuum chamber, we use an avalanche photodiode (APD) with an $852~\mathrm{nm}$ bandpass filter, placed outside of the chamber, to collect the scattered light from the chip components. The APD position is adjusted to maximize the collected scattered signal. This APD signal is then used for active stabilization of the trap power.

\section{Electronic control system}
\label{app.electronics}

We use ARTIQ 7 (Advanced Real-Time Infrastructure for Quantum physics) as the experimental control system~\cite{ARTIQ2016}. The control hardware is based on a Kasli-SoC FPGA and includes a Zotino DAC card, three Urukul DDS cards, two DIO cards, and three Sampler ADC cards. Microwave drive signals at $6.5~\mathrm{GHz}$ are generated using a Valon 5015 synthesizer. A $10~\mathrm{MHz}$ $^{87}$Rb frequency standard (FE-5680A) serves as the common reference clock for the entire experimental system (See Fig.~\ref{fig.appendix_figure3}(h)).

\begin{figure}[!t]
\includegraphics[width=.95 \columnwidth]{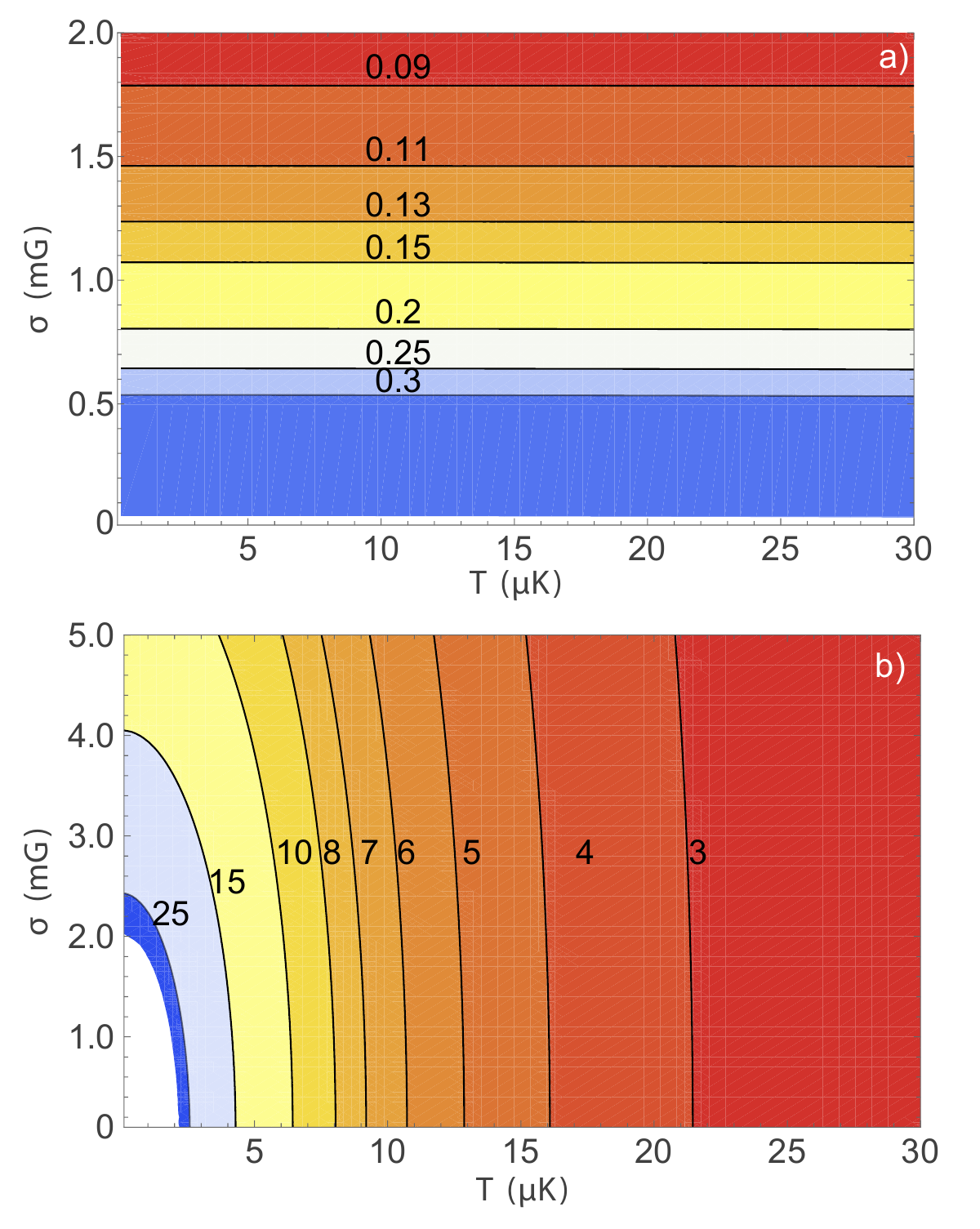}
 \caption{\rsub{Calculated Ramsey coherence time as a function of atom temperature and magnetic field noise at a bias field of $B_0=3.23~\rm G$ and a trap wavelength of 852 nm.  We assume  Gaussian magnetic field noise with standard deviation of $\sigma$. Contours are labeled with the coherence time in ms. a) Calculated coherence of the $\ket{1,1}\rightarrow\ket{2,1}$ transition which is primarily sensitive to magnetic noise. The observed coherence time of $T_2^*=0.11~\rm  ms$ indicates $\sigma=1.45~\rm mG$. b) Calculated coherence of the $\ket{1,0}\rightarrow\ket{2,0}$ clock transition. The observed coherence time of $T_2^*=3.32 ~\rm ms$ indicates a temperature of $T=20~\mu\rm K$.}}
 \label{fig.appendix_fig_AtomT}
\end{figure}

The two-photon transition $\ket{\downarrow} \leftrightarrow \ket{\uparrow'}$ is driven by simultaneous application of a microwave tone at $\sim~6.832~\mathrm{GHz}$ and an RF tone at $2.5~\mathrm{MHz}$. The RF field is generated by a loop antenna consisting of four turns of wire with a diameter of $7~\mathrm{cm}$, placed directly on top of the vacuum chamber as close as possible to the atom. The antenna is not impedance matched to the output of the RF amplifier (Mini-Circuits LZY-22$+$), so only a small fraction of the available RF power is converted into RF radiation.

\begin{table*}[!t]
\footnotesize
\caption{\rsub{Representative demonstrations of  ion-photon entanglement. The success probability is defined as the probability of detecting a photon per attempted ion excitation and the rate is the ion-photon entanglement generation and detection rate. Reported fidelities are measured results without correction for separately characterized error sources.}}
\label{tab.ion_photon_entangle}
\centering
\begin{tabular}{|l| l| l| c| c| c| c| }
\hline
Year &  Description & Photon encoding & Matter qubit & Success probability & Rate $(\rm s^{-1})$ & Fidelity \\
\hline
2004 \cite{Blinov2004}  Monroe &  lens, $NA=0.23$ & polarization & $^{111}$Cd$^+$  &$1.6\times10^{-4}$ & $0.3~$ &$0.87~$\\
2012 \cite{Stute2012}  Blatt &  cavity, $C=1.75$  & polarization & $^{40}$Ca$^+$  & $5.7\times10^{-2}$ & $40.5~$ &$0.974\pm0.002~$\\
2018 \cite{Bock2018}  Eschner &  lens, $NA=0.4$  & polarization& $^{40}$Ca$^+$  & $4.76\times10^{-4}$ (a) & $27.6~$ (b)&$0.933 \pm 0.003~$ (c)\\
2020 \cite{Stephenson2020}  Lucas  & lens, $NA=0.6$  &  polarization & $^{88}$Sr$^+$  & $2.08\times10^{-2}~$ (d) & $4000 (5700)~$ & $0.979(0.977)~$ \\
2021 \cite{Kobel2021}  Kohl  & cavity, $C=0.056$  &  polarization & $^{171}$Yb$^+$  &$2.5\times10^{-3}$ & $62~$ &$0.901 \pm 0.017~$\\
2024  \cite{OReilly2025} Monroe  &  lens, $NA=0.8$ &  polarization  & $^{138}$Ba$^+$  & $2.31 (2.21) \times10^{-2}~$  & $7692(7359)~$ (e) & $ 0.981 (0.968)~$\\
2025   \cite{Saha2025} Monroe&   lens, $NA=0.6 (0.8)$ & time bin & $^{138}$Ba$^+$  & $4.75(9.5)\times10^{-3}~$ (f) &  $71 (142)~$ &
$0.985~$ (g) \\
\hline
\end{tabular}\\
\rsub{ 
(a) Calculated from reported entanglement generation and detection rate of 27.6/s with 58 kHz sequence repetition rate. 
(b) The paper reports $236~$ generated and $27.6~$ projected and detected entanglement events/s with repetition rate of $58~$kHz.
(c) With background subtraction, the paper reports $0.955\pm0.003~$.
(d) Calculated from ion-ion entanglement success probability of $2.18 \times 10^{-4}$.
(e) Calculated from reported attempt rate of $333~$ kHz.
(f) Calculated from reported values in Method section of \cite{Saha2025}.
(g) We list the square root of the ion-ion entanglement fidelity reported in \cite{Saha2025}.
}
\end{table*} 
\normalsize

\section{\rsub{Atom temperature measurement}}
\label{app.atomTemp}
\rsub{The temperature of the trapped atoms is calculated based on the coherence time of the clock transition from the Ramsey data shown in Fig.~\ref{fig.figure2}(e). We calculated the Ramsey coherence as a function of both atomic temperature and magnetic field noise. The results are shown as a two-dimensional color map in Fig.~\ref{fig.appendix_fig_AtomT}. Following Ref.~\cite{Kuhr2005}, the calculation is based on thermal averaging of the differential light shift experienced by the atom in the trap, together with the residual sensitivity of the clock transition to magnetic-field fluctuations \cite{Saffman2011}. For each temperature, the atomic motion is treated within the thermal distribution of the trapping potential, which gives a distribution of light shifts and therefore a corresponding inhomogeneous dephasing of the Ramsey signal. Magnetic field noise is included as an additional Gaussian broadening of the transition frequency. The coherence of the mapping transition (see Fig. \ref{fig.figure2}f) primarily depends on the magnetic field 
which implies noise of $\sigma=1.45~\rm mG$.
Since the clock qubit at our bias field is only weakly sensitive to magnetic field fluctuations, the calculated coherence time depends predominantly on the atomic temperature over the relevant parameter range. By comparing the measured value $T_2^* = 3.32~\mathrm{ms}$ to the calculated map, we infer an atomic temperature of approximately $20~\mu\mathrm{K}$.}

\section{\rsub{Ion-photon entanglement}}
\rsub{Although the focus of this work is atom–photon entanglement with neutral atoms, we include in Table~\ref{tab.ion_photon_entangle} a summary of representative ion–photon entanglement experiments for comparison and reference. The substantially higher rates and fidelities achieved with trapped ions should be viewed in the context of important platform-dependent differences as well as experimental and fundamental constraints. }

\bibliography{qc_refs,saffman_refs,atomic,optics,rydberg, Additional_refs}

\end{document}